\documentclass[english,prl,twocolumn]{revtex4}
\usepackage[T1]{fontenc}
\usepackage{amsmath,hyperref,amssymb}
\usepackage{graphicx,url}
\usepackage{float}
\usepackage{subfigure}

\begin{document}

\title{Phase transitions of magnetic AdS$_4$ black holes with scalar hair}

\author{Kiril Hristov$^{\dag}$, Chiara Toldo$^*$, Stefan Vandoren$^*$}
\affiliation{* Institute for Theoretical Physics \emph{and} Spinoza Institute,
 Utrecht University, 3508 TD Utrecht, The Netherlands,}
 
 \affiliation{\dag Dipartimento di Fisica, Universit\`a di Milano-Bicocca,
I-20126 Milano, Italy, and \\ INFN, sezione di Milano--Bicocca, I-20126 Milano, Italy}

\begin{center}
\begin{abstract} 
We determine the thermodynamic properties of a class of spherically symmetric and static black holes in AdS$_4$ with magnetic charges and scalar hair. These black holes are solutions in 4d $N=2$ gauged supergravity that can arise from eleven-dimensional supergravity compactified on S$^7$. At zero temperature, they preserve supersymmetry and hence are stable. At non-zero temperatures, we explore in detail the canonical ensemble and stability of solutions and find a first order phase transition between small and large hairy black holes. The transition emerges as a liquid-gas phase transition in the dual three-dimensional field theory on $\mathbb{R}\times $S$^2$ with magnetic flux through S$^2$. 
\end{abstract}
\end{center}

\maketitle

\section{Introduction and summary}\label{sect:intro}

Black holes behave much like thermodynamical systems. They have an internal energy (the mass), a temperature (due to Hawking radiation) and other quantum numbers like charges and angular momentum. The entropy is determined by the area of the event horizon \cite{Bekenstein:1973ur,Hawking:1974sw}, and from these data one can compute the free energy. Black holes can undergo phase transitions, which reveal a particularly interesting structure in anti-de Sitter spacetimes (AdS). Most famous is perhaps the Hawking-Page phase transition between a thermal gas of gravitons and Schwarzschild black holes in AdS  \cite{Hawking:1982dh}. Using the AdS/CFT correspondence, this phase transition was interpreted as the transition from a confining to a de-confining phase in the dual gauge theory \cite{Witten:1998zw}. Similarly, phase transitions of charged black branes (black holes with planar symmetry) with scalar hair have been used in modeling holographic superconductors on the boundary, where a charged condensate forms via a second order phase transitions and a global U(1) symmetry is spontaneously broken \cite{Hartnoll:2008vx}.

In this work we present a new phase transition for spherically symmetric black holes, based on a class of magnetic solutions with non-trivial scalar profiles. These magnetic black holes were recently discovered in \cite{Toldo:2012ec,Klemm:2012yg} and can be thought of as thermal excitations over supersymmetric (BPS) static black holes with fixed graviphoton magnetic charge \cite{Cacciatori:2009iz,Dall'Agata:2010gj,Hristov:2010ri}. Our analysis follows closely the study of charged AdS$_4$ black holes presented in \cite{Chamblin:1999tk,Cvetic:1999ne,Chamblin:1999hg}, but it is applied to a different class of black hole solutions. The phase transition we find shows nevertheless some similarities to the one found in \cite{Chamblin:1999tk,Chamblin:1999hg} for electrically charged black holes and in the absence of scalar fields \footnote{The black holes studied in \cite{Chamblin:1999tk,Chamblin:1999hg} and \cite{Duff:1999gh} can also be magnetic, but are then related to the electric ones by electro-magnetic duality. Our solutions are not of that kind and have to be studied separately.}.

\begin{figure}[H]
\centering
{\includegraphics[width=70mm]{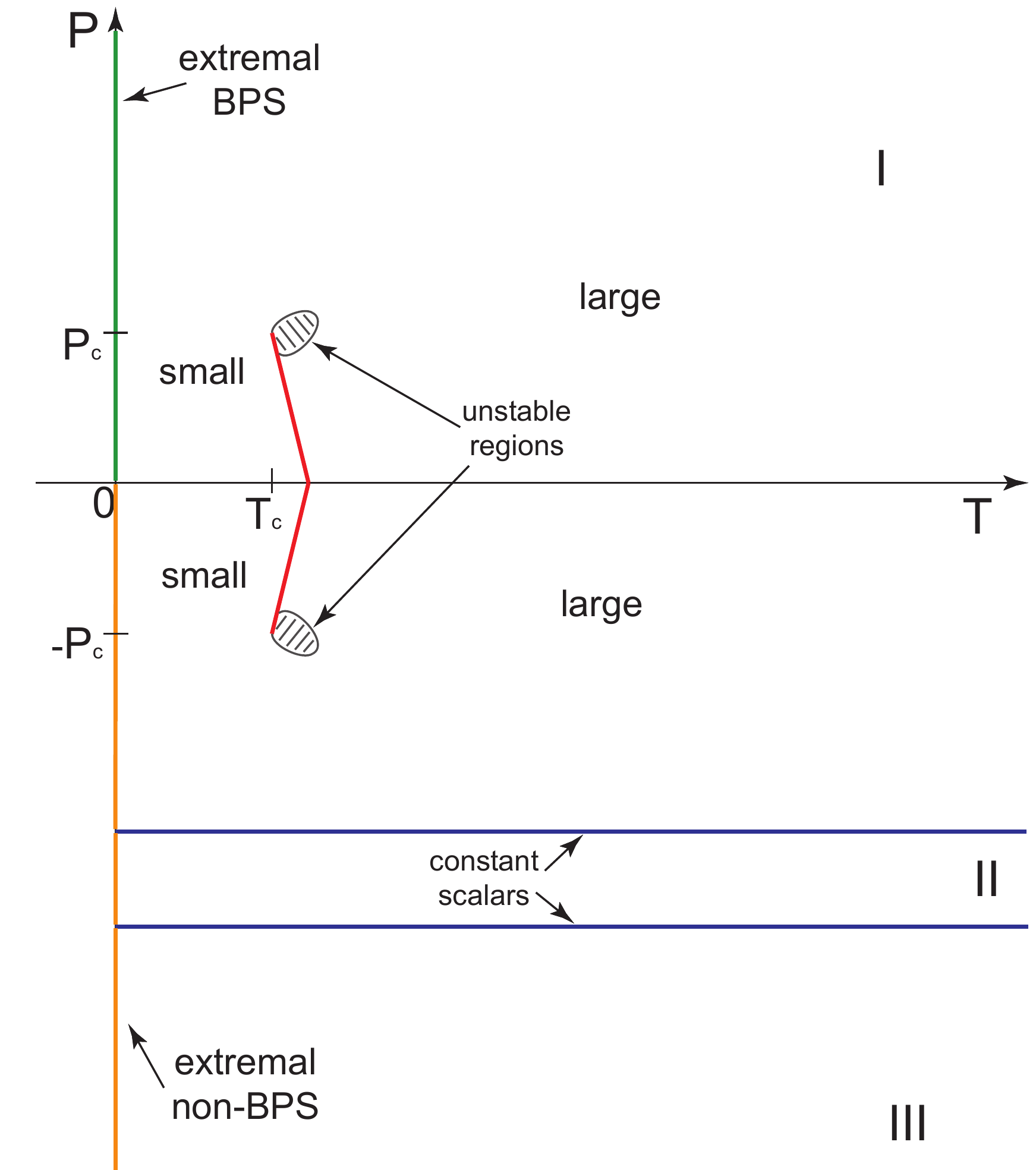}}
\caption{{\footnotesize The phase diagram in the canonical ensemble. Regions I, II and III are separated by the two blue lines of thermal black holes with constant scalar profiles.}}
\label{fig1}
\end{figure}

Anticipating our final results, we already present the phase diagram in the canonical ensemble in Fig.\ \ref{fig1}, the main result of our study. We denote by $P=p^1$ one of the two magnetic charges carried by the black hole. In region I we find a small/large black hole first order phase transition for $|P|<P_c$ and a smooth crossover between the two phases for $|P|>P_c$. 
Apart from a very small region close to criticality, the phase diagram consists of solutions with positive specific heat that are stable against thermal fluctuations. In regions II and III, as well as the two lines with thermal black holes for constant scalars, BPS solutions are absent, and one finds a smooth crossover between small and large black holes that are thermally stable everywhere in phase space \footnote{For constant scalars, the magnetic charge is fixed by the Dirac quantization condition and falls outside the critical region. Hence there is no phase transition in our case for constant scalars.}. 

Figure 1 needs to be supplemented with the phase diagram for the scalar hair, which we present in the last section of the paper.
There, we also discuss the implications of this result for the AdS/CFT correspondence.

\section{Black hole solutions}\label{sect:solutions}
We will discuss properties of black hole solutions of gauged 4d $N=2$ supergravity \cite{DeWit:1984px,Andrianopoli:1996cm}. 
We consider the particular case of the $N=2$ gravitational multiplet coupled to a single vector multiplet  with gauge coupling constant $g$ and prepotential $F = - 2 i \sqrt{X^0 (X^1)^3}$. The gauge group is  U(1)$\times$U(1), and the gauging introduces charges for the two gravitini and a scalar potential $V$ that contains Fayet-Iliopoulos (FI) parameters. The Lagrangian is written in e.g. \cite{Toldo:2012ec}, whose conventions and notation we follow here. This is the simplest four-dimensional theory that allows  for static supersymmetric black holes in AdS$_4$ with spherical symmetry. Moreover, for certain values of the FI parameters, the model can be directly embedded in eleven dimensional supergravity \cite{Duff:1999gh,Hristov:2010ri}, hence this could lead to an underlying microscopic string/M-theory description.

The FI parameters of the theory, denoted by $\xi_0$ and $\xi_1$, determine the electric charges for the two gravitini \footnote{One can also consider magnetic gaugings and obtain dyonic BPS and non-BPS configurations, like in \cite{Dall'Agata:2010gj,Gnecchi:2012kb}.} (with respect to the $U(1)$ gauge fields $A_{\mu}^0$ and $A_{\mu}^1$). They are $e_\Lambda\equiv g\xi_\Lambda$ ($\Lambda=0,1$) for one gravitino and opposite for the other. The magnetic charges $p^0$ and $p^1$ of the black hole therefore obey a Dirac quantization condition,
\begin{equation}\label{quantization}
	g \xi_{\Lambda} p^{\Lambda} = n\ , \qquad n \in \mathbb{Z}\ .
\end{equation} 
In this work we choose $n=-1$ (the case of $n=+1$ is then also covered by a sign flip on the gravitino charges), which ensures that the black holes asymptote to the stable magnetic AdS ground state \cite{Hristov:2011ye}. For values of $n \neq \pm 1$, spherical black holes do not preserve supersymmetry at infinity, so we expect instabilities to occur.

The magnetically charged black hole background solutions of  \cite{Toldo:2012ec} have all fermions set to zero and are given by the spacetime metric 
\begin{equation}\label{ansatz_metric}
{\rm d}s^2= U^2(r) {\rm d}t^2 - U^{-2}(r) {\rm d}r^2- h^2(r) ( {\rm d}\theta^2+ \sin^2 \theta {\rm d}\varphi^2)\,,
\end{equation}
with
\begin{equation}\label{nonextr_new_sol}
U^2(r)= e^{\mathcal{K}}\left(g^2 r^2+c - \frac{\mu}{r}+\frac{Q}{r^2} \right)\,, \qquad  h^2(r) = e^{-\mathcal{K}} r^2 \,.
\end{equation}
In these coordinates, the field strengths are given by $F^\Lambda=p^\Lambda \sin\theta\, {\rm d}\theta \wedge {\rm d}\varphi$ and we do not allow for electric charges. The vector multiplet contains a complex scalar field $z$ that parametrizes a special K\"ahler manifold with 
K\"ahler potential $\mathcal{K}$ given by
\begin{equation}
z = \frac{X^1}{X^0}\ , \qquad \mathcal{K}=-log[X^0 \overline{X}^0 (\sqrt{z}+\sqrt{\bar{z}})^3]\,.
\end{equation}
The general thermal solutions with running scalars have
\begin{equation}\label{c1}
X^0 = \frac{1}{4 \xi_0} - \frac{\xi_1 b_1}{\xi_0 r}\ , \quad X^1 = \frac{3}{4 \xi_1} + \frac{b_1}{r}\ , \quad c = 1 -\frac{32 (g \xi_1 b_1)^2}{3} \ ,
\end{equation}
\begin{equation}\label{c2}
Q= -\frac{16}{3} b_1^2 \xi_1^2 + \frac{1}{g^2} -\frac{256}{27} b_1^4 \xi_1^4 g^2 + \frac{2 \xi_1 p^1}{g} +\frac{4}{3} \xi_1^2 (p^1)^2\,,
\end{equation}
and
\begin{equation}\label{mu}
\mu=\frac{8}{3} \xi_1 b_1  - \frac{3}{4 g^2 \xi_1 b_1} +\frac{512}{27}g^2 \xi_1^3 b_1^3  - \frac{3 p^1}{2 g b_1} -\frac{2\xi_1 (p^1)^2}{3 b_1}\,.
\end{equation}
The configuration (\ref{ansatz_metric}-\ref{mu}) is a solution to the full non-linear set of coupled equations of motion for the metric, gauge fields and complex scalar. For fixed gravitino charges $e_\Lambda$, the solution contains two parameters, which we can choose to be $b_1$ and $p^1$ (with $p^0$ fixed by the quantization condition \eqref{quantization}).

The solution for the scalar field is real. Its asymptotic expansion is
%\begin{equation}
$z=3\frac{\xi_0}{\xi_1}+16\xi_0\frac{b_1}{r}+{\cal O}(r^{-2})$
%\end{equation}
where the constant term is dictated by the value of the scalar in the (magnetic) AdS$_4$ vacuum at infinity. The extremum of the potential $V_*$ sets the value of the cosmological constant which in our conventions is $\Lambda=3 g^2 V_*=-2{\sqrt 3}g^2{\sqrt {\xi_0\xi_1^3}}$ \cite{Toldo:2012ec}. After truncating the imaginary part of $z$, we can write down the Lagrangian for the real part, canonically normalized as $\phi={\sqrt{\frac{3}{8}}}\ln z$. We then get, using the results of \cite[App.\ 1]{Toldo:2012ec},
\begin{eqnarray}
L&=&\frac{1}{2}R(g)-e^{\sqrt{6}\phi} F^0_{\mu\nu}F^{0\,\mu\nu}-3e^{-\sqrt{\frac{2}{3}}\phi} F^1_{\mu\nu}F^{1\,\mu\nu}\nonumber\\
&+&\frac{1}{2}\partial_\mu\phi \partial^\mu\phi +g^2\large(\xi_0\xi_1e^{-\sqrt{\frac{2}{3}}\phi}+\frac{\xi_1^2}{3}e^{\sqrt{\frac{2}{3}}\phi}\large)\ .
\end{eqnarray}
For values for which $3\xi_0=\xi_1$, one gets the cosh-potential of \cite{Duff:1999gh}, obtained from a truncation of eleven-dimensional supergravity compactified on S$^7$. The extremum of the potential is at $\phi_*=\sqrt{\frac{3}{8}}\ln(3\xi_0/\xi_1)$, which is the maximum. Expanding around this value, one finds the mass of the scalar to be $m_{\phi}^2=-2g^2|V_*|/3$. One can then verify that the mass satisfies the Breitenlohner-Freedman bound, and in the conventions of e.g. \cite{Hertog:2004ns} in which $V_*=-3$, we have $m_{BF}^2=-9/4<m_{\phi}^2=-2<m_{BF}^2+1$. Asymptotically, the scalar field behaves as
\begin{equation}\label{phi-asympt}
\phi-\phi_*=\frac{\alpha}{r}+\frac{\beta}{r^2}+{\cal{O}}(r^{-3})\ ,
\end{equation}
with $\beta =\alpha^2/{\sqrt{6}}$ and $\alpha=4\sqrt{2/3}\,b_1\xi_1$. We have, again in the conventions of \cite{Hertog:2004ns},
$\beta=k\alpha^2$ with a fixed value of $k=1/{\sqrt{6}}$ \footnote{Usually, scalar hair solutions allow for arbitrary values of $k$, but it might be that the absence of electric charges and the coupling of the scalars to the gauge fields, have fixed this parameter to a particular value. It would be interesting to study dyonic extensions of our black hole solutions in which this parameter is free.}.

The magnetic charge $P\equiv p^1$ can be varied freely and defines three regions of black hole solutions with running scalars: $P >P_I$, $P_I > P > P_{III}$ and $P <P_{III}$. The values $P_{I} = -\frac{3}{4 g \xi_1}$ and $P_{III} = -\frac{3}{2 g \xi_1}$ correspond to the only smooth solutions with constant scalars ($b_1=0$), see \cite{Toldo:2012ec}. For other values of the magnetic charge, the limit to constant scalars $b_1=0$ is not smooth. This is an important difference compared with the situation in \cite{Chamblin:1999tk,Chamblin:1999hg}, where there are no scalars and the magnetic charge is free.

The temperature of the black hole can be calculated from the surface gravity, and is given by $T=\frac{1}{2\pi}|UU'|_{r=r_+}$, or
\begin{equation}\label{temp}
T =\frac{1}{4 \pi} \bigg{|} e^{\mathcal{K}(r_{+})} \bigg( 4 g^2 r_{+} - \frac{\mu}{r_{+}^2} +2 \frac{c}{r_{+}} \bigg) \bigg{|}\,,
\end{equation}
with the value of the radius at the outer horizon $r_+$ found as the largest root of the equation $U^2(r) = 0$. The resulting Bekenstein-Hawking entropy is therefore
\begin{equation}\label{entropy}
S=\frac{A}{4} = \pi h^2(r_+) = \pi r_+^2 e^{-\mathcal{K} (r_+)}\ .
\end{equation}
\begin{figure}[H]
\centering
{\includegraphics[width=57mm]{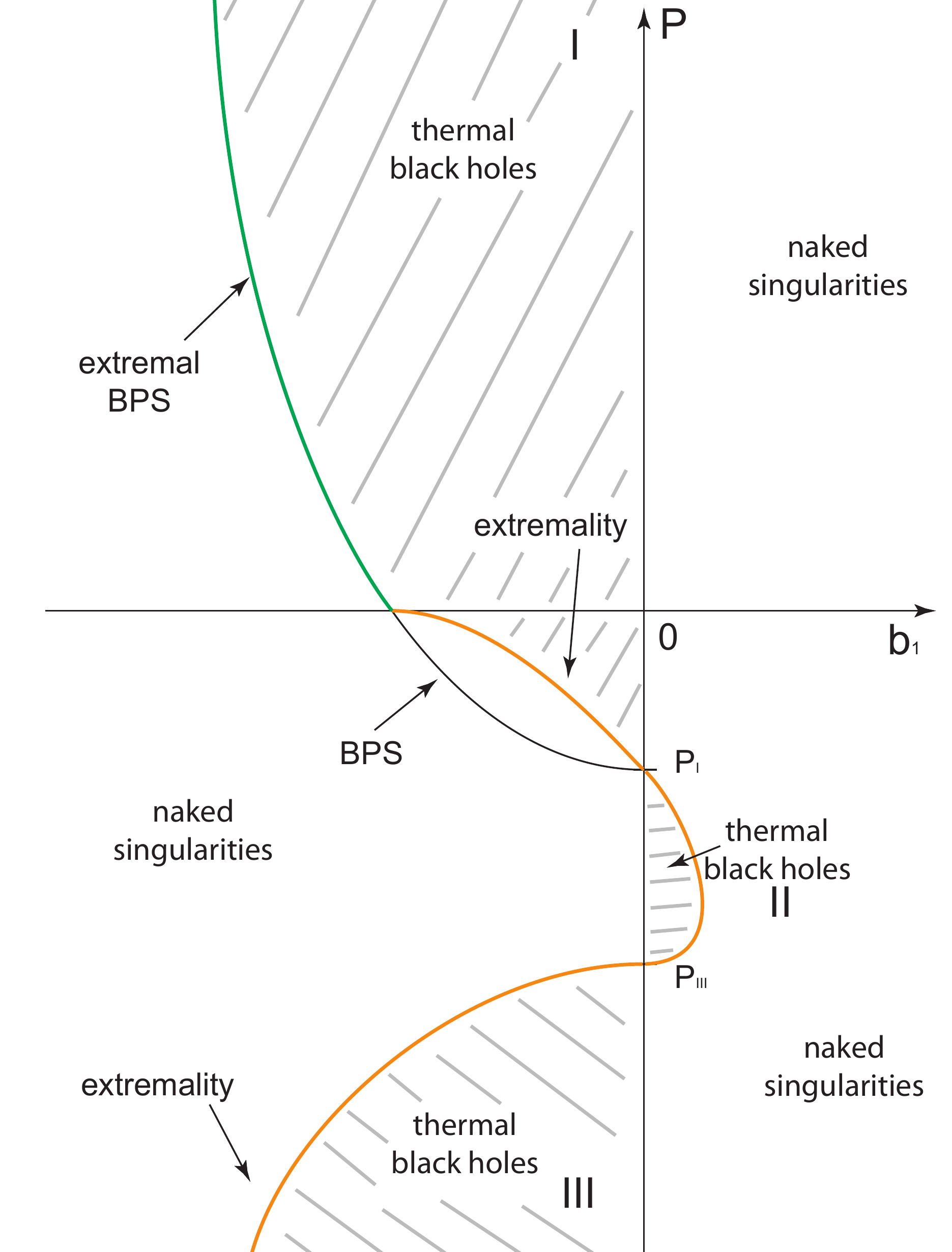}}
\caption{{\footnotesize Parameter space for genuine black hole solutions (shaded) in regions I, II and III.}}
\label{fig2}
\end{figure}
A black hole solution can therefore be plotted in the $(p^1, b_1)$ plane, as shown in Fig. \ref{fig2}. Not all points correspond to solutions with horizons. Singular and regular solutions are separated by two curves. The first one (green curve) corresponds to the 1/4-BPS solutions of \cite{Cacciatori:2009iz,Dall'Agata:2010gj,Hristov:2010ri}, and is defined by the relation $p^1=-\frac{1}{4e_1}(3-2\alpha^2g^2)$, such that $U^{2}(r)=e^{K}(gr+\frac{c}{2gr})^2$ with horizon located at $r_+^2=-c/(2g^2)$ with $c<0$.

The second curve (in orange) defines the extremal, but non-BPS black hole solutions. It can be obtained from setting $T=U^2=0$, excluding the BPS black holes and solutions with naked singularities. It yields conditions that can be analyzed numerically and the result is in Fig. \ref{fig2}.

\section{Thermodynamics and phase transitions}\label{sect:ensembles}
%Having defined the thermal solutions of interest, we can now in principle evaluate the Euclidean 4d supergravity action for the black holes, subtract the appropriate ground state contribution and find rigorously the thermodynamic potentials in the canonical and grand-canonical ensemble. This is the path taken in previous literature \cite{Chamblin:1999tk,Cvetic:1999ne,Chamblin:1999hg}, which we decide to shortcut here by looking at the first law of thermodynamics. 

In this section we discuss the thermodynamical properties of the black holes in the canonical ensemble. For magnetically charged black holes, the grand-canonical ensemble seems less appealing, since there are no magnetically charged particles the black hole can emit. One can also allow the cosmological constant $\Lambda$ to vary, as done in \cite{Cvetic:2010jb,Kubiznak:2012wp}. We however keep $\Lambda$ fixed and we set $V_*=-3$. In particular we set $\xi_1=3/\sqrt2$, $\xi_0=1/\sqrt2$, which allows for an embedding into eleven-dimensional supergravity, as discussed before.

Black holes with magnetic charges must satisfy the first law of thermodynamics
\begin{equation}\label{first_law}
	{\rm d} M = T {\rm d} S + \chi_{\Lambda} {\rm d} p^{\Lambda}\ ,
\end{equation}
where the magnetostatic potentails are defined as $\chi_{\Lambda}=-\int_{r_+}^{\infty} G_{\Lambda,tr} {\rm d}r$, and $G$ is the dual field strength. For the case at hand, these potentials can be computed to be
\begin{equation}
\chi_0= -\frac{1}{\sqrt2} \frac{\sqrt2+3 p^1}{(\sqrt2 r_+ -12 b_1)} \,, \qquad \chi_1= \frac{1}{\sqrt2} \frac{3 p^1}{(\sqrt2  r_+ +4b_1)}\,.
\end{equation}
All quantities on the right hand side of \eqref{first_law} have now been determined, so we can integrate the expression to find the mass $M$ on the left hand side,
\begin{equation}\label{energy}
	M= 2 \sqrt2 b_1  -\frac{1}{4 \sqrt2 \, g^2 b_1} -\frac{3p^1}{4 g b_1}-  \frac{ (p^1)^2}{ \sqrt2 b_1}\,.
\end{equation}
This expression for the mass can be reproduced by using the AMD-formalism \cite{Ashtekar:1984zz,Ashtekar:1999jx}, which we explicitly verified. The AMD mass in asymptotically AdS space guarantees that $M$ is a conserved quantity\footnote{Surprisingly, this quantity does not coincide with the one coming from the superalgebra, computed in \cite{Hristov:2011qr,Hristov:2011ye,Toldo:2012ec}.}.

%\subsection{Canonical ensemble}
The canonical ensemble describes a closed system at fixed temperature and charge, ${\rm d} T = {\rm d} p^\Lambda = 0$. The free energy
\begin{equation}\label{free_energy}
	F = M - T S\ ,
\end{equation}
is extremised at equilibrium, ${\rm d} F = 0$, due to the first law \eqref{first_law}. One can therefore compare the free energy $F(T,P)$ of different solutions at fixed $T$ and $P\equiv p^1$ - the state with lowest $F$ is the one that is thermodynamically preferred.

%\subsection{Grand-canonical ensemble}
%The grand-canonical ensemble allows for charge changes in the system, keeping the magnetic potentials $\chi_0, \chi_1$ fixed, ${\rm d} \chi_{\Lambda} = 0$. The grand-canonical potential can be defined as
%\begin{equation}\label{gr_can}
%	W = M - T S - \chi_{\Lambda}  p^{\Lambda}\ ,
%\end{equation}
%and is also extremised due to the first law, ${\rm d} W = 0$. $W$ is the Legendre transform of $F$, i.e.\ $F = W + \chi_{\Lambda}  p^{\Lambda}$. This ensemble is relevant when black holes can dynamically change their magnetic charges, which happens in the process of Hawking radiation of charged particles. Such magnetically charged particles, however, are not present in 4d $N=2$ gauged supergravity, unless the black hole can emit smaller black holes with magnetic charge. Our analysis therefore mostly focusses  on the canonical ensemble.
%\section{Phase structure and stability}\label{sect:phases}

\begin{figure}[H]
\centering
{\includegraphics[height=15mm]{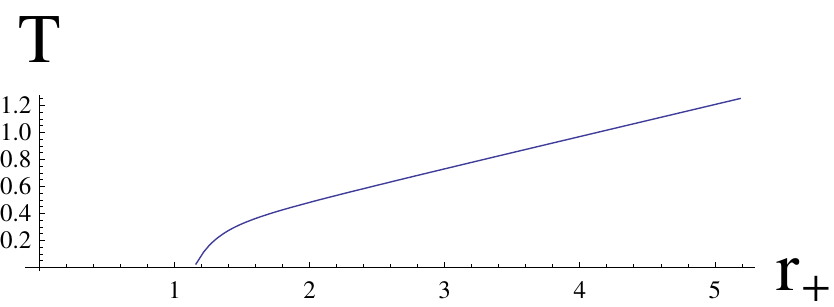}}
{\includegraphics[height=15mm]{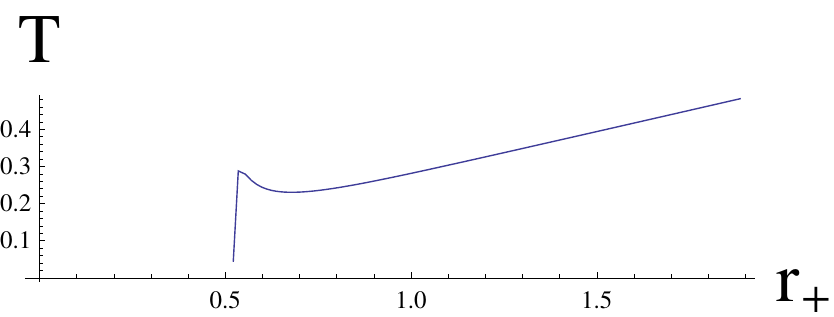}}
\caption{{\footnotesize Temperature vs.\ outer horizon radius $r_+$ for fixed values of $|P|>P_c$ (left, $P=0.5$) and $|P|<P_c$ (right, $P=0.01$). Numerical analysis shows that $P_c\approx 0.025$ for the values $\xi_1=3/\sqrt2$, $\xi_0=1/\sqrt2$ and $g=1$.}}
\label{fig3}
\end{figure}

The phase diagram in the canonical ensemble was already presented in Fig.\ \ref{fig1}, but now we can discuss it in more detail. It is instructive to compare how the temperature behaves as a function of the outer horizon radius $r_+$ for constant magnetic charge in the region $-P_{c} \leq P \leq P_{c}$ and outside it (Fig.\ \ref{fig3}).
For $|P|>P_c$, the temperature is a monotonic function of the radius, and so there exists only one black hole for fixed $T$. In the  region $|P|<P_c$, for fixed $T$ there exist up to three different black holes - a small, a medium-sized, and a large black hole. 
To see which black hole is preferred in the canonical ensemble, we plot the free energy for these solutions: 
\begin{figure}[H]
\centering
{\includegraphics[height=30mm]{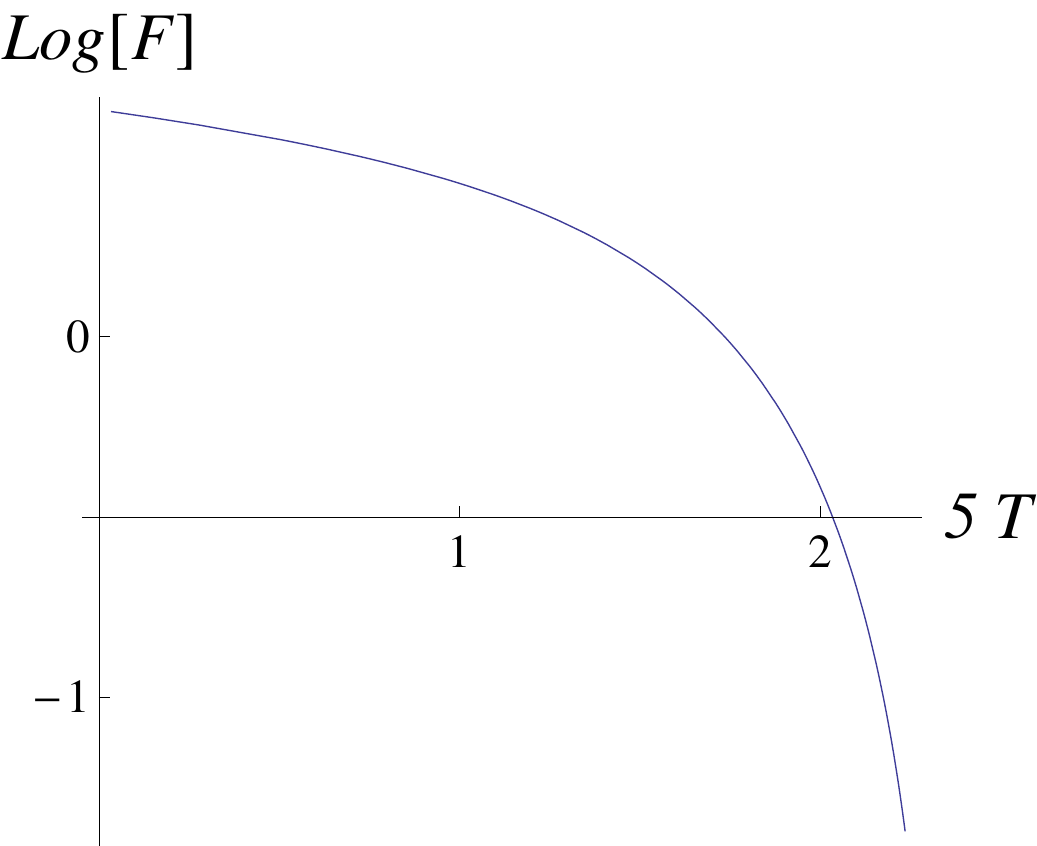}}
{\includegraphics[height=30mm]{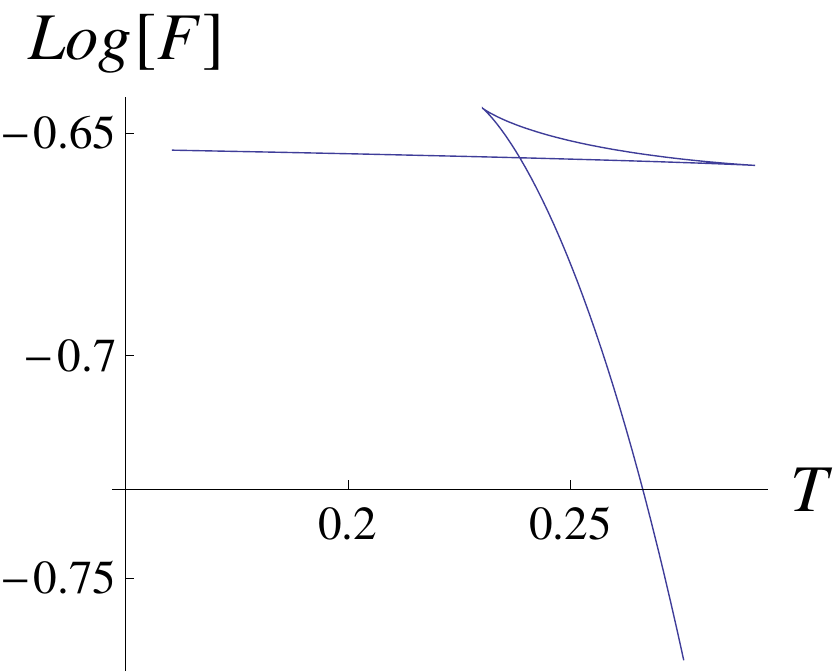}}
\caption{{\footnotesize Free energy as a function of temperature for the same fixed values of $P$ as in Fig.\ \ref{fig3}.}}
\label{fig4}
\end{figure}
 From Fig.\ \ref{fig4} we see that for $|P|<P_c$, increasing $T$ leads to the appearance of medium-sized and large black hole branches, whose free energy is initially higher than for small black holes \footnote{We have an exception of this only at $P=0$, where the branch with larger black holes dominates over the smaller black holes everywhere. Therefore the critical lines do not cross the $P=0$ axis, just as in \cite{Chamblin:1999tk,Chamblin:1999hg}.}. Upon further increase of $T$, the free energy of the large black holes rapidly decreases, leading to a crossing point. The free energy is not smooth here, and so a first order phase transition between small and large black holes occurs, similar to \cite{Chamblin:1999tk,Chamblin:1999hg}.

To check the intrinsic stability of the different phases against thermal fluctuations, one computes the specific heat $C_P$ at fixed magnetic charge, 
\begin{equation}
	C_P \equiv T \left( \frac{\partial S}{\partial T} \right)_P \geq 0\ ,
\end{equation}
which must be positive for stable solutions. It turns out that the medium-sized black holes have negative specific heat and are therefore unstable under temperature fluctuations. The other regions on Fig.\ \ref{fig3}, where the temperature grows with size, have a positive specific heat and are stable. This means that both the small and the large black hole phases are stable. There is a tiny region where this is not exactly true, as shown on Fig.\ \ref{fig1} - close to the critical temperature $T_c\approx 0.226$, the phase with larger size black holes contains some negative specific heat solutions according to our numerical analysis. The exact values and sizes of each feature on the phase diagram depend on the values of the free positive constants $g, \xi_0, \xi_1$. One finds, for instance, that the value of  $P_{c}$ scales inversely with the coupling constant $g$, and we expect $P_c \sim (-\Lambda)^{-1/2}$, and $T_c \sim (-\Lambda)^{1/2}$, with cosmological constant $\Lambda$.

\section{Dual description via AdS/CFT}\label{sect:conclusion}

The supergravity models we have discussed can be embedded in M-theory compactified on AdS$_4\times$ S$^7$ \cite{Hristov:2010ri,Duff:1999gh}, and will have field theory duals in the class of ABJM models \cite{Aharony:2008ug} defined on $\mathbb{R}\times $S$^2$. The bulk gauge group U(1)$\times$U(1) corresponds to part of the global R-symmetry group on the boundary, which can be gauged to produce background magnetic fields on the boundary with flux through the S$^2$. The bulk model contains a scalar field with $m_\phi^2=-2$, so the boundary values provide sources for operators of dimensions equal to one or two, depending on the choice of quantization. Examples of such operators in three dimensions are bilinears of boundary scalars $\varphi$ or fermions $\psi$ transforming under the global R-symmetry group are 
\begin{equation}
{\cal O}_1={\rm Tr}(\varphi^I a_{IJ}\varphi^J)\ ,\qquad {\cal O}_2={\rm{Tr}}(\psi^I b_{IJ}\psi^J)\ ,
\end{equation}
for some constant matrices $a$ and $b$. According to the holographic dictionary, the expectation values of these operators are given by the boundary values of the bulk scalar field, $\langle {\cal O}_1 \rangle =\alpha$ and $\langle {\cal O}_2 \rangle =\beta$. Moreover, the ABJM action $S_0$ is deformed by multitrace operators, $S=S_0+k\int {\cal O}_1^3$, along the lines of \cite{Witten:2001ua,Berkooz:2002ug,Hertog:2004ns}. The dimensionless deformation parameter $k$ is given by the boundary conditions of the scalar field, $\beta=k\alpha^2$, using the notation of (\ref{phi-asympt}). In our case, it is fixed to be $k=1/{\sqrt 6}$.

Since $\alpha \sim b_1$ and $\beta \sim b_1^2$, we can use $b_1$ as an order parameter and express it as a function of the temperature. Doing so, we get the following plots,

\begin{figure}[H]
\centering
{\includegraphics[height=20mm]{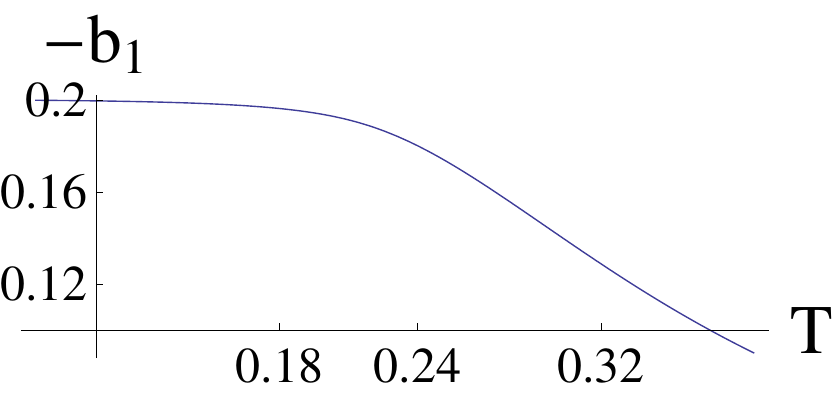}}
{\includegraphics[height=22mm]{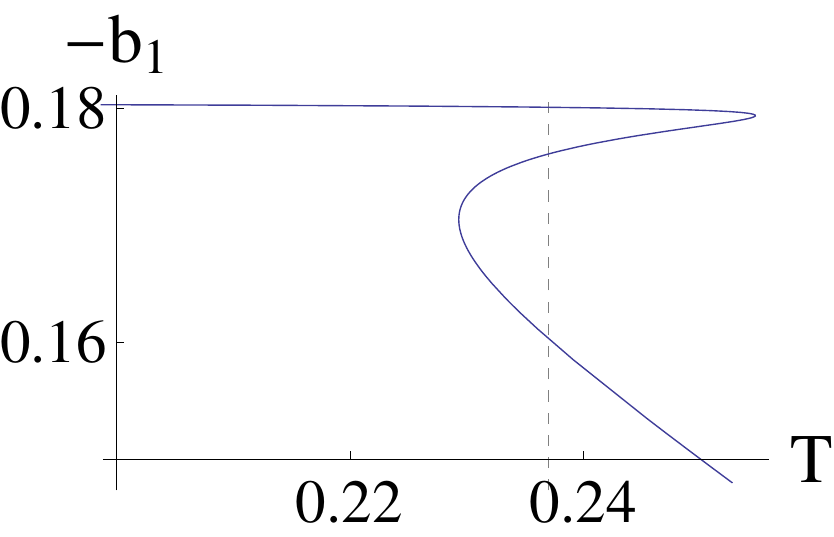}}
\caption{{\footnotesize Behaviour of the order parameter $b_1$ in function of the temperature $T$ for values $P>P_{c}$ (left, $P=0.1$) and $P<P_{c}$ (right, $P=0.015$).}}
\label{fig7}
\end{figure}

The condensate is actually never vanishing for any finite temperature, and it goes to zero in the limit $T \rightarrow \infty $, so the global R-symmetry is always broken. The behavior of the order parameter mimics more that of a liquid-gas transition, and  one can determine the critical temperature from  Maxwell's 'equal area' construction.

The thermodynamical behavior of the black holes discussed in this paper should be reproduced by the boundary field theory thermal partition function  on S$^1\times $ S$^2$. At zero temperature, degeneracies of the ground state of this system should explain the finite entropy of the BPS-black hole, presumably through the presence of Landau levels that arise  because of the magnetic field. At finite temperature, we predict that the system undergoes a phase transition for magnetic charges below a critical value. It would be interesting to understand this phase transition directly in the boundary field theory, perhaps using the techniques recently developed by e.g. \cite{Jain:2013py}.

\section*{Acknowledgements}
We thank J. de Boer, A.\ Gnecchi, S. Minwalla, M. Rangamani, S. Ross, H. Stoof and A. Zaffaroni for valuable discussions. C.T.\ and S.V.\ acknowledge support by the Netherlands Organization for Scientific Research (NWO) under the VICI grant 680-47-603. K.H.\ is supported in part by INFN and by the MIUR-FIRB grant RBFR10QS5J ``String Theory and Fundamental Interactions''.

\end{document}